\documentclass{owrart}

\usepackage{color}

\usepackage{amsthm}
\newtheorem{theorem}{Theorem} 

\begin{document}

%% --------------------------------------------------------------------------
%% Please use the environment "talk" for each abstract.
%% It has three obligatory and one optional argument. The syntax is:
%% -----------------------
%% \begin{talk}[coauthors]{Name of the speaker}{Title of the talk}{Author Sorting Index}
%%      .....
%% \end{talk}
%% -----------------------
%% The names of coauthors will appear in form of "(joint work with ...)"
%%
%% The Author Sorting Index should be given as the last and first name of the speaker,
%% separated by a comma. If for example the name of the speaker is "John Smith", then
%% the correct Author Sorting Index is "Smith, John".
%% Any special characters (like accents, German umlaute, etc.) should be replaced by
%% their "non-special" version, eg replace \"a by a, \'a by a, etc.
%%
%% Please use the standard thebibliography environment to include
%% your references, and try to use labels for the bibitems, which
%% are uniquely assigned to you in order to avoid conflicts with other authors.
%% You can achieve unique labels by using our on initials before every label.
%% -------------------------------------------------------------------------------

\begin{talk}[]{Daniela Cadamuro}
{Quantum backflow in scattering situations}
{Cadamuro, Daniela}

\noindent

In classical (statistical) mechanics, point masses travel in the same direction as their momentum. 
For a particle moving without friction in one direction, its initial state is a probability distribution $\sigma(x,p)$, where $x,p$ are position and momentum of the particle, with time evolution $\sigma_t(x,p) = \sigma(x-pt, p)$. The probability density $\rho_t(x) = \int dp\; \sigma_t(x,p)$ to find the particle at a point $x$ and the corresponding probability flux $j_t(x) = \int dp\; p\, \sigma_t(x,p)$ fulfill the continuity equation, implying
\begin{equation}\tag{$\ast$}\label{flux}
\frac{d}{dt} \int_0^\infty dx\; \rho_t(x) = j_t(0).
\end{equation}
We note that if $\sigma_t(x,p)=0$ for $p<0$ and all $x$, then $j_t \geq 0$, i.e., the right hand side of \eqref{flux} is non-negative. Therefore, the probability of finding the particle on the right of $0$ is increasing in time, while it is decreasing on the left. In other words, if the particle has positive momentum (with probability 1) then the probability distribution moves to the right.

For a quantum mechanical particle the situation is different. Suppose that its state $\varphi$ is chosen so that it still has momentum $p > 0$ with probability 1. Can we still expect $\langle J (x) \rangle_\varphi \geq 0$ (where $\langle J(x) \rangle_\varphi$ denotes the expectation value of the probability flux operator $J(x)$ in the state $\varphi$)?
In some examples for $\varphi$, exactly this does \emph{not} happen, which is called ``\emph{quantum backflow}''. The question is how large this effect is, and whether positivity is retained in some approximate sense.

As a first example, we consider a \emph{free} particle (without external forces) moving in one dimension, whose state is a vector in the Hilbert space $\mathcal{H}= L^2(\mathbb{R},dp)$ with support in $\mathbb{R}_+$ (positive momentum with probability 1). Its time evolution is given by $\varphi_t(p) = \exp(-i\frac{p^2}{2m}t)\varphi(p)$. The probability density and probability flux for the position of the particle are
\begin{eqnarray*}
\langle \rho(x) \rangle_\varphi &=& \int dp \,dq\; \overline{\varphi(p)} \frac{1}{2\pi} e^{i(q-p)x} \varphi(q),\\
\langle J(x) \rangle_\varphi &=& \int dp \,dq\; \overline{\varphi(q)}\frac{p+q}{4\pi} e^{i(q-p)x} \varphi(q)
\end{eqnarray*}
and the continuity equation still holds. We can find states with positive momentum such that the expectation value of the flux operator at a given point $x$ is negative, see \cite{EFV05} for explicit examples. Hence, the probability of finding the particle is increasing in time to the left of $x$, and decreasing to the right (``backflow'').
To compute lower bounds on the flux, some care is needed. As the expression above diverges for large momenta $p,q$, it makes sense only as a quadratic form. We therefore consider ``smooth averages'' over a small space interval: $\langle J_f \rangle_\varphi = \int dx\; f(x) \langle J(x) \rangle_\varphi$ with $f \in \mathcal{S}(\mathbb{R})$, $f \geq 0$.

We now ask about the spectrum of $J_f$ restricted to wave functions of positive momentum $\varphi = E_+ \varphi$ (where $E_+$ is the projector onto positive momenta). 
The following result is due to Eveson, Fewster and Verch \cite{EFV05}:
\begin{theorem}
For every $f \in\mathcal{S}(\mathbb{R})$, $f \geq 0$, there exists $c_f \geq 0$ such that $\langle J_f \rangle_\varphi \geq -c_f  \| \varphi \|^2$ whenever $\varphi$ has compact support in $\mathbb{R}_+$.
\end{theorem}
This ``quantum inequality'' shows that the backflow effect is limited. However, the result does not take into account any influence from external potentials. Our goal is to prove that quantum backflow is a \emph{general} feature of quantum mechanical particles, even if they are not free. Can one expect such lower bounds in the presence of a potential? There may be a reflected wave, which behaves (for large times) like a free particle moving to the left -- is this an obstacle to bounds?

We consider a scattering potential $V$ and the Hamiltonian $H = \frac{1}{2m}P^2 + V(x)$. The time evolution is now more intricate since it is not just a multiplication operator. However, for suitable potentials, the wavefunctions have a well-known asymptotics at $t \rightarrow \pm \infty$ (incoming/outgoing waves), which is an essential ingredient of \emph{scattering theory}. This is of importance since a \emph{prima facie} problem to the extension of the result \cite{EFV05} in the interacting situation is to understand what a particle ``moving to the right'' is. In fact, since $[E_+, H] \neq 0$, the space of ``positive momentum wavefunctions'' is not preserved under time evolution.

A replacement for this notion is provided in scattering theory by considering asymptotic (incoming) states with positive momentum. The M\o ller operator $\Omega_{\mathrm{in}}$ maps ``interacting'' states to ``incoming'' states. Using this, the question we ask is whether $E_+ \Omega_{\mathrm{in}}J_f \Omega_{\mathrm{in}}^\ast E_+$ is bounded below. This is investigated in a joint work
with H.~Bostelmann and G.~Lechner \cite{BCL17} and our main result is:
\begin{theorem}
Suppose that $\int dx\; (1 + |x|) \lvert V(x) \rvert < \infty $. Then, for any $f \in \mathcal{S}(\mathbb{R})$, $f \geq 0$, there exists a constant $c_{V,f} >0$ such that
\begin{equation*}
E_+ \Omega_{\mathrm{in}} J_f \Omega_{\mathrm{in}}^\ast E_+ \geq -c_{V,f} \cdot \pmb{1}.
\end{equation*}
\end{theorem}
This applies to all usual short range potentials; examples are the square well potential (attractive or repulsive), the P\"oschl-Teller type transparent potentials, and any measurable potential decaying like $|x|^{-\alpha}$, $\alpha >2$. A similar result holds for a delta potential (attractive or repulsive). This indicates that reflection does not present a problem to existence of lower bounds.

This result could potentially be generalized to other, physically interesting situations, including particles with spin or other degrees of freedom, several particles, 2- or 3-dimensional scattering. One could also investigate possible generalizations of the current $J$, for example consider the probability flux of only the ``spin up'' component in the case of a spin-$1/2$ particle, or identify a suitable $J$ in a 3-dimensional situation. Another possible direction for generalization is to consider PDEs other than the Schr\"odinger equation, or more generally, recasting the theorem in the framework of abstract scattering theory. Finally, we remark that backflow may be one of the quantum mechanical effects directly verifiable in experiments, as an experiment to measure it has been proposed \cite{PTMM13}. Moreover, it has relations to other ``quantum inequalities'' appearing in quantum field theory in connection with the energy density, which are relevant for the stability of spacetime; see, e.g., \cite{F12,BC16}.

\end{talk}

\end{document}